# Hot Electron Bolometer Development for a Submillimeter Heterodyne Array Camera

Matthew O. Reese, Daniel F. Santavicca, Luigi Frunzio, Daniel E. Prober

*Abstract*—We are developing Nb diffusion-cooled Hot Electron Bolometers (HEBs) for a large-format array submillimeter camera. We have fabricated Nb HEBs using a new angle deposition process. We have characterized these devices using heterodyne mixing at 20 GHz. We also report on optimizations in the fabrication process that improve device performance.

*Index Terms*—Hot Electron Bolometers, superconductivity proximity effect, electron-beam fabrication

## I. Introduction

A superconducting microbridge connected to thick, normal metal contacts forms the active element of the superconducting Hot Electron Bolometer (HEB), which has demonstrated promising performance as a terahertz detector.[1-5] HEBs have several desirable characteristics: unlike SIS tunnel junctions, they are not limited by the gap frequency;[6] they require very small local oscillator power; their simple geometry, with low stray impedance, facilitates integration in multi-pixel arrays; and they have demonstrated IF bandwidth as large as 9 GHz.[1-5]

Recently, we have focused on fabricating diffusion-cooled niobium HEBs. These devices are being developed to be part of a large format array camera for use in the 810 GHz atmospheric window on the Heinrich Hertz Telescope operated by the University of Arizona. This can serve as a model for future THz camera designs. We require a reasonably sharp resistive transition of the superconductor, a critical temperature ($T_c$) of 4-5 K to operate in a pumped $^4$He cryostat, and contact pads that either do not superconduct or have a much lower $T_c$ than the bridge. Here we present our new fabrication process along with initial characterization results.

## II. Fabrication Method

The essential goal has been to produce Nb HEBs using a single lithographic patterning and no cleaning step between

Manuscript received October 19, 2005. This work was supported by NSF-AST, NASA-JPL, and a NASA Graduate Student Research Fellowship for M. O. Reese.

All authors are in the Department of Applied Physics of Yale University, New Haven, CT 06520-8284 USA. All correspondence should be directed to Daniel Prober (phone: 203-432-4280; fax: 203-432-4283; e-mail: daniel.prober@ yale.edu).

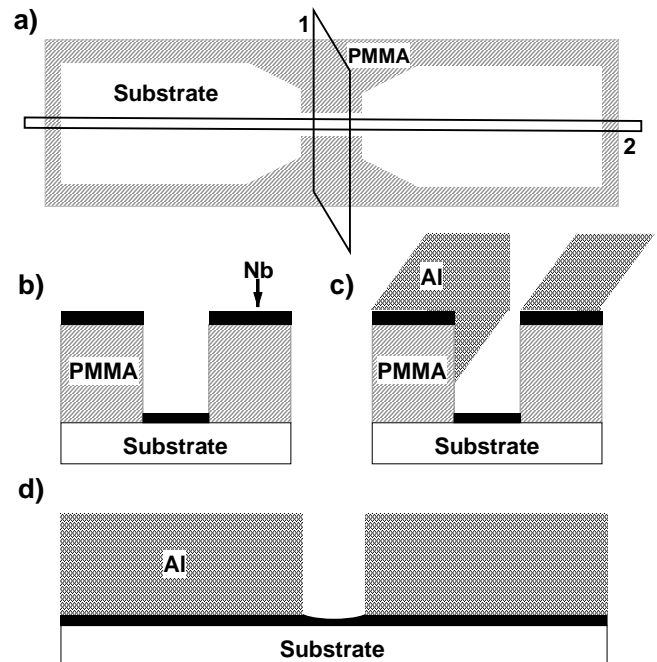

Fig. 1. Deposition process: a) e-beam pattern PMMA, top view [Side views of slice 1: Fig. b&c, slice 2: Fig. d], b) sputter Nb, c) angle evaporate Al, Al sticks on side of resist in bridge region, d) final result after liftoff. The Nb in the bridge center is slightly thinner than the contact region (see text).

the deposition of our Nb and normal metal. (Here we use Al, because $T_{bath} > T_{c,Al}$). Our structure is patterned as shown in Fig. 1. We use a converted Scanning Electron Microscope (FEI Sirion XL40) to expose a pattern in a 380 nm thick monolayer of 950K polymethyl methacrylate (PMMA) spun on a substrate of silicon or fused silica. [When using fused silica, we evaporate 10-15 nm of Al on top of our PMMA before e-beam writing, to reduce charging effects. Before developing our resist, we remove the Al with 5 minutes in MF-312 developer (4.9% by volume tetra methyl ammonium hydroxide in water), then arrest the process with 20 s in isopropyl alcohol (IPA).] Developing is done in a solution at 25°C of 1:3 methyl isobutyl ketone:IPA for 20 s in an ultrasonic bath, arrested by 20 s in IPA in ultrasound. The substrate is then blown dry with $N_2$. The sample is loaded into a Kurt J. Lesker Supersystem III Series multi-sputtering and evaporation system with a base pressure of ~$10^{-6}$ Pa. First the 2" diameter Nb target is pre-sputtered for 2 min in an Ar plasma with a power of 350 W and a pressure of 0.17 Pa and a flow rate of ~82 sccm. The sample is then ion beam cleaned by Ar with current of 4.7 mA for 15 s using a 3 cm Kaufman-type gun to help adhesion to the substrate. Then, a Nb film is sputtered for 9 s, using the same parameters as stated above. And then with as little delay as possible (typically ~3 min.), we angle evaporate 200 nm of Al at 1 nm/s. Lift-off is done in hot acetone (50°C) for one hour, followed by 2 min in ultrasound.

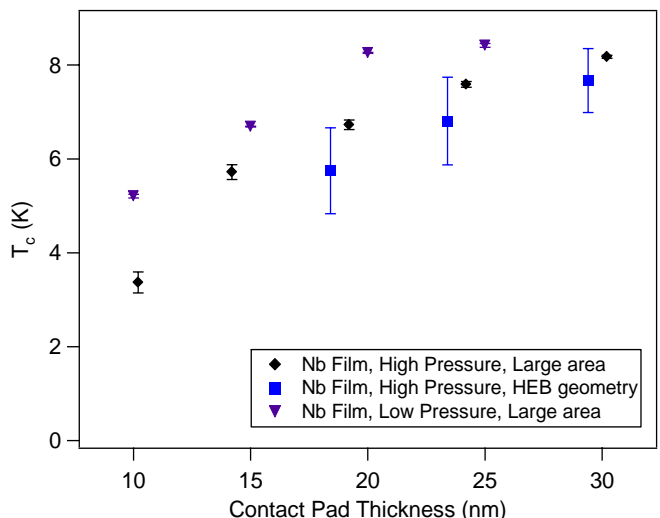

Fig 2. Transition temperature of Nb Films with no Al on top. The error bars represent the onset and completion of the transition ($T_c$ is defined as the center of the transition). The points are spread out horizontally to allow easier viewing, and suggest the error in thickness measurements. The thicknesses are 10, 14, 19, 24, and 30nm, measured in the contact pad region.

|  | Estimated Nb Thickness | Base Pressure | $\Delta t$: Nb$\rightarrow$Al |
|---|---|---|---|
| HEB A | 10 nm | $4 \times 10^{-5}$ Pa | 35 min |
| HEB B | 10 nm | $4 \times 10^{-5}$ Pa | 6 min |
| HEB C | 12 nm | $1 \times 10^{-5}$ Pa | 37 s |

Table I. Fabrication parameters for the Hot Electron Bolometers appearing in Fig. 3.

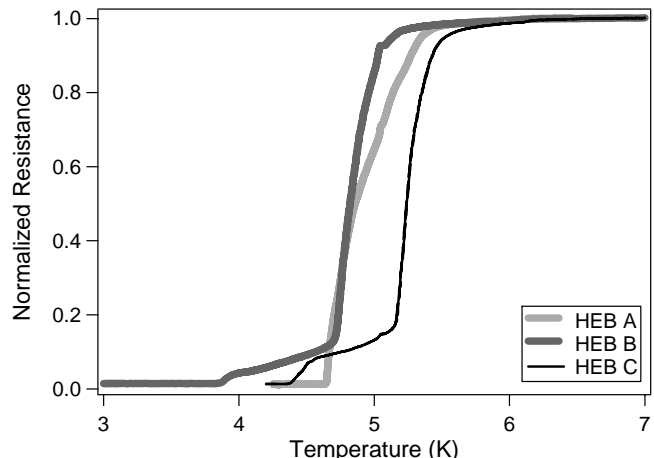

Fig 3. Normalized resistance versus temperature curves for three HEBs. See Table for details on fabrication parameters. Notice the foot structure for devices B and C due to the contact pads. The broad resistive transition of Device A may be a result of poor suppression in the contact pads as well.

### III. RESULTS AND OPTIMIZATIONS

#### A. Monolayer Resist

Our fabrication development originally used a bilayer resist process (PMMA on top of the copolymer MMA) designed to allow clean liftoff. This works well with sputtered Nb and even with the directionality of evaporated Al films. However, while liftoff was clean, our sputtered Nb HEBs had broad resistive transitions. This resulted from the non-directional nature of sputtering. The Nb spread out to the base of the bottom resist layer, which had undercut of 50-100 nm. Hence, the center of a 200 nm wide bridge was only ~60% of the thickness of the contact pads, from spreading of the Nb after it had passed through the narrow opening in the top resist layer.

Using a single layer of resist significantly reduced the effect of Nb spreading. As the bridge becomes increasingly narrow (<1 μm), however, reduction of the bridge thickness is still observed. This can be seen in Fig. 2 by the broadening of the resistive transitions for narrow bridges. This effect is due to the sputter process and the geometry of the PMMA slot used to form the HEB. Sputtering is a non-directional process. In a large opening in the resist pattern, sputtered material from every direction will reach the substrate. In a narrow opening, a fraction of the incident Nb will not have a line of sight to the substrate, thereby reducing the deposited thickness in the center of the wire compared to the contact pad region. This reduced thickness leads to a broader and lower temperature transition for the bridge compared to the contact pads.

#### B. Improved Nb

We were able to improve the quality of our thin film Nb, as determined by its $T_c$, by lowering our Ar plasma pressure and at the same time increasing the dc power [See Fig. 2]. Our original values were 0.53 Pa/270 W; our new values are 0.17 Pa/350 W. This increased our deposition rate from 1.0 nm/s to 1.25 nm/s and may have reduced the film stress and thus improved the quality.[7]

#### C. Suppression of Superconductivity in the Contact Pads

We want to form a clean interface in our contact pads, thereby ensuring good suppression of the superconductivity in the thin Nb under the thick Al. The presence of a superconducting energy gap in the pads would inhibit the ability of hot electrons to diffuse off the bridge. Because the proximity effect has a strong dependence on both the thickness of the Nb and Al films as well as the transparency of the interface between them, it can be difficult to achieve good suppression.[8,9] We found that the thicknesses of our Nb and Al films, as well as the length of time between their depositions, significantly affect suppression.

We first achieved good suppression in the contact pads (while still having an acceptably high $T_c$ in the bridge) by reducing the time delay between depositions. Our improved Nb films have increased $T_c$ and reduced $\Delta T_c$. This allows for sufficiently thin (10 nm) Nb films, which are more readily suppressed. We also use a thick (200nm) normal metal film. By reducing the base pressure of our system from $4 \times 10^{-5}$ to $1 \times 10^{-6}$ Pa, we were able to make the time interval between depositions relatively unimportant. With P = $10^{-6}$ Pa we could get the same suppression if we waited 30 s or 30 min. With our higher quality Nb film parameters and 200 nm Al, at a base pressure of 4 x $10^{-5}$ Pa, we made two batches of devices with a long (HEB A) and shorter (HEB B) time interval between the Nb and Al depositions to illustrate the importance of interface transparency [Fig 3].

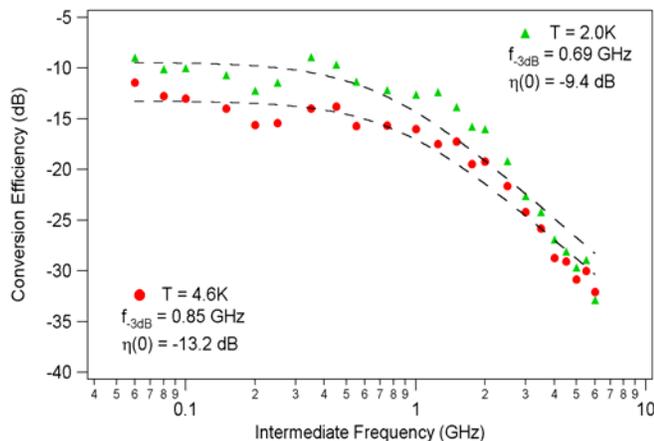

Fig 4. IF Conversion efficiency of HEB C with bandwidth above and below the transition of the suppressed contact pads. The contact pads in this sample had a $T_c$=4.4K.

HEB A has a single, somewhat spread out transition. The contact pads have a higher onset transition temperature than the bridge, because the contacts pads are thicker than the bridge, and only weakly suppressed. HEBs B and C each had short intervals between depositions and each has two distinct transitions. The higher temperature transition is the bridge transition; the second is the transition of the suppressed contact pads. The transition temperature of the contact pads (~3.8 K for HEB B, ~4.4K for HEB C which had thicker Nb) agrees with that of Nb-Al bilayers, made during the same deposition but measured separately. The finite resistance remaining after the bridge transition is in part due to the large number of squares in the rf choke structure in our device geometry, but largely a result of the ends of our bridge being proximitized normal.[10,11]

### D. Improved Bandwidth

To illustrate the importance of good suppression of the superconductivity of the contact pads, we provide bandwidth measurements in the overpumped regime, where the critical current is suppressed by the local oscillator power. These were performed at temperatures above and below the $T_c$ of the contact pads. These data are from HEB C, a 400 nm long device, from the same cooldown [Fig. 4]. These results were obtained by heterodyne mixing at 20 GHz following the techniques described in Ref. 9. We used wirebonds, however, instead of a flip-chip process to connect the device. The difference in conversion efficiencies between the two curves can be explained by the typical increase of conversion efficiency as the bath temperature is lowered to $T_{bath} \approx T_c/2$.[12] There is an observed increase of ~15% of the IF bandwidth when the contact pads are normal, at T=4.6 K. The reduced bandwidth at T=2 K may be due to hot electrons at energies less than the gap of the contact pads that cannot easily diffuse off the nearly normal bridge. This would increase the thermal response time and thus reduce the IF bandwidth. At 4.6 K, the gap in the contacts is zero and all the hot electrons are able to leave the bridge rapidly. The bandwidth, to the authors' knowledge, has not previously been observed to exhibit any temperature dependence in a diffusion-cooled device. A 2 GHz bandwidth is desirable, and can readily be achieved by reducing the bridge length to about 250 nm.[13]

### IV. FUTURE PLANS

In the future, we seek to reduce the $T_c$ of the contact pads beyond our present reduction, thereby allowing us to operate at 2 K and maximize both conversion efficiency and bandwidth. We also plan to investigate the mixer noise and conversion efficiency at 345 GHz, in measurements at the University of Arizona.

### ACKNOWLEDGMENT

We thank Bertrand.Reulet, Lafe Spietz, and John D. Teufel for helpful discussions.

### REFERENCES

[1] E. M. Gershenzon, et. al., Sov. Phys. Supercond. **3**, 1990, 1582
[2] D. E. Prober, Appl. Phys. Lett. **62**, 1993, 2119
[3] J. J. A. Baselmans et. al., Appl. Phys Lett. **84,** 2004, 1958
[4] R. A. Wyss,, et. al., Proc. 10th Intl. Symp. Space Terahertz Tech, 1999, 215
[5] B. S. Karasik, et. al., IEEE Trans. Appl. Supercond. **7**, 1997, 3580
[6] M. Bin, et. al., IEEE Trans. Appl. Supercond. **7**, 1997, 3584
[7] S. Knappe, C. Elster, H. Koch, J. Vac. Sci. Technol. A **15,** 1997, 2158
[8] J. M. Martinis, et. al., Nucl. Instrum. Methods A **444**, 2000, 23
[9] D. Esteve, et. al, "Correlated fermions and transport in mesoscopic systems", eds. T. Martin, G. Montambaux, and J. T. T. Van (1996)
[10] D.W. Floet, et. al., Appl. Phys. Lett. **73**, 1998, 2826
[11] I. Siddiqi, et. al, J. Appl. Phys. **91**, 2002, 4646
[12] P.J. Burke, Ph.D Thesis, Yale University, 1998, accessible from http://www.yale.edu/proberlab/alumni.html
[13] P. J. Burke, et. al., Appl. Phys. Lett. **68**, 1996, 3344